# HARVESTING AND STORING LASER IRRADIATION ENERGY WITH GRAPHENE-CU COMPOUND STRUCTURE

## ABSTRACT

Graphene-metal compound structure has been reported as a novel and outstanding component used in electrical and optical devices. We report on a first-principles study of graphene-cu compound structure, showing its capacity of converting laser energy into electrical power and storing the harvested energy for a long time. A real-time and real-space time-dependent density functional method (TDDFT) is applied for the simulation of electrons dynamics and energy absorption. The laser-induced charge transfer from copper layer to graphene layer is observed and represented by plane-averaged electron difference and dipoles. The effects of laser frequency on the excitation energy and charge transfer are studied as well. The enhancement of C-C σ-bond and decreasing of electron density corresponding to π-bond within graphene layer are responsible for the ability of storing the harvested energy for a long time.

## INTRODUCTION

At present, graphene is considered to be an excellent candidate for development of carbon-based electronics and optoelectronics, thanks to its properties, such as very high electron mobility, high current-carrying density and excellent heat dissipation.[1-3] Resent progress in converting solar and mechanical energy into electric power by using carbon based materials has opened a door for a brand-new way of energy harvesting with high conversion efficiency.[4-7] In addition carbon-based supercapacitor shows its outstanding property of high values of capacitance and energy density[8, 9] which indicates the possibility of storing electrical charges efficiently. Of particular interest is the fact that graphene-based materials have demonstrated controllable surface and interface properties as well as tailored work functions via functionalization during synthesis and/or post-treatment. Since the graphene can be n-type or p-type doped through contacting with different metals[10], it inspires us the fascinating applications of harvesting irradiation energies such as laser. In despite of previous works which have studied the electronic transport through metal-graphene junction and the global photo-response in epitaxial graphene with metal contacts,[6, 7, 11, 12] it is advantageous to explore the full picture of the electronic dynamics of the contacts at the nanometer scale with the existence of laser.

Here, we present a TDDFT study of the electronic dynamics of the graphene-Cu compound structure subjected to external short laser pulse. It is found that electron transfer from copper layer to graphene sheet occurs when the compound structure is under the laser irradiation. Duo to the electronic oscillation of graphene sheet, the external applied lasers with different frequency result in different electron transfers. The in plane σ-bond of graphene enhances after the laser irradiation while π-bond decreases its electronic density. We infer this is responsible for the long-time life storage of the harvested energy based on this compound structure. A plane capacitor model could be used to describe this structure, then harvesting and storing laser irradiation energy is the procedure of charging the capacitor.

## COMPUTATIONAL DETAILS

Density functional theory (DFT) calculations are performed in our simulation. The interfacial structure is built using orthogonal (111) fcc surface unit cells containing two layers of copper with a graphene sheet adsorbed on top of the metallic slab and a vacuum region of 25 Å. The irreducible wedge of the Brillouin zone is presented by 224 k points to represent the Bloch wave functions for the momentum-space integration. DFT pseudopotentials are used for the core electrons of both carbon and copper atoms, interactions between valence electrons and ions are treated by the norm-conserving pseudopotentials with separable nonlocal operators.[13] The electron wave functions are expanded in a plane-wave basis with an energy cutoff of 37 Ry.

The structure configuration of graphene-Cu interface was first optimized. The generalized gradient approximation (GGA)[14] is used to study the surficial geometry here. A key issue in the calculation is accommodating the 3.8% mismatch between Cu (111) surface and graphene. In despite of variety of schemes for lattice mismatch accommodation employed in different DFT simulations,[15-17] it is reasonable to set the in-plane lattice constant to that of metal with the total energy conversing to minimum.[18] For this reason, we chose the in-plane lattice constant equal to that of Cu (111) in our simulation, adjusting graphene lattice constant accordingly. Although previous publications have pointed out that there are four possible interface geometries,[16, 17, 19, 20] our optimized configuration comes out to be the low-energy and stable one (shown in figure 1). The graphene-Cu interlayer distance is 3.42 Å – in agreement with previous work.[10, 18] As the consequence of the weak interaction, the graphene and copper layers are almost decoupled from each other.

Since the charge transfer does not depend strongly on the choice of density functional when the graphene-Cu structure is correctly obtained,[16] a hybrid exchange-correlation functional – PEBh[21] is used during the time evolution simulated by using the OCTOPUS code.[22] The TDDFT calculation is based on the adiabatic density approximation. We represent the laser irradiation by subjecting our system to an external alternating electric field parallel to the layers – x-direction in our model. The applied laser pulses of 21 a.u. (≈ 0.5 fs) duration are Gaussian wave packets with the same peak power density $5 \times 10^{14}$ W/cm$^2$ but different photonic energy of 100 eV and 500 eV respectively, as shown in figure 2. During the electron dynamic simulation, the time step is 0.01 a.u. and the ions are frozen.

## RESULTS AND DISCUSSION

The graphene-Cu compound structure does not absorb the same energy under the irradiation of different lasers with different frequencies despite the identical power density. The excitation energies as a function of time as well as the electric field of applied laser pulses are shown in figure 2. The excitation energies change slightly at the initial stage of the laser pulses where the applied electric fields are still weak, but it has an abrupt change after 8 a.u. (≈ 0.2 fs). After the laser pulses are applied, the excitation energies keep unchanged because there are no more energy exchanges with the outside environment. It is found that the laser with higher frequency, the excitation energy absorbed is higher. There are about 1.29 Ha (about 35 eV) difference between the excitation energies after the irradiation of the two kinds of laser pulses.

One might expect that the absorption of excitation energies should have some connection with the number of electrons transferred; it is natively to guess that the higher the excitation energy is, the more electrons will be transferred. But the following analysis shows the reality is not that simple. It is known that dipoles along the axes can reflect the charge transfer in different directions. Therefore, we report the dipoles along the three axes as a function of time shown in figure 3. The signals of dipole along x axis follow the laser profiles with the existence of the external alternating electric fields, and oscillate around their initial states after the laser pulses. Only a slight disturbance could be observed for the dipoles along y axis no matter with or without the external alternating electric fields. Quite different to the signals along x axis and y axis, the dipoles along z axis do not oscillate around the initial states after the irradiation. They increase quickly under the irradiation of laser pulses and slightly decrease to final states instead of the initial ones when the electric fields decay to zero. The results suggest that there exists charge transfer along z axis which is perpendicular to the graphene sheet and copper layer. The insert in figure 3(a) shows the dipole along z axis for a long time. It still remains oscillating even at the time of 150 a.u. (≈ 3.7 fs) with no tendency to decaying to the initial state which means that the graphene sheet of the system have the ability of holding the charges transferred from the copper layers. In this way, the compound structure harvests the laser energy and converts it into electric power to store for a relatively long time.

In order to characterize the charge transfer between the interfaces, we can use the plane-averaged electron densities $n(z)$ to visualize the electron redistribution upon irradiation of laser pulses. $\Delta n(z)=n_t(z)-n_i(z)$, where $n_t(z)$ and $n_i(z)$ denote the plane-averaged densities of system states at time $t$ and origin, respectively. The transferred-charge $q$ (per carbon atom) can be estimated by integrating $\Delta n(z)$ from the node between the metal surface and the graphene sheet.[16] From figure 4 we can see the changes of $q$ with time. The charge transfer begins as soon as the laser pulses are applied and reaches the maximum during the irradiation time, and then begin to decease. At very last of our simulation, the transferred-charge also oscillates slightly around a definite value. Although the more energy the system absorb from laser fields, the fewer charges transfer between the graphene sheet and copper layer. The charge $q$ induced by the laser pulse with photonic energy of 100 eV is about 0.06 at most and 0.04 at last, as compared with 0.03 and 0.02 respectively when the photonic energy increases to 500 eV. The changes of $q$ with time are consistent with the changes of dipoles along z axis with time. In order to understand this phenomenon, we calculate the electronic oscillation spectrum of the graphene sheet. The existence of a peak around 100.5 eV indicates the resonance absorption is the main reason which causes the phenomenon. This result is consistent with the experimental work which shows that the laser pulses with different wavelength induce different photocurrent maxima at the same pulse energy but the photocurrent has the same order of magnitude.[6, 7]

Figure 5 shows the electron density difference between the states before and after the irradiation of laser pulse in the x = 0 plane and the graphene sheet plane as well. The electron redistribution as shown in figure 5(a) between Cu (111) surface and graphene sheet shows the electron decrease of copper layer and the increase of graphene. Although the charge transfers from copper to graphene, the π-bond of graphene losses its electron density after the irradiation which is presented by green around the carbon atom in the figure, unexpectedly. In contrary to the π-bond, the in-plane C-C σ-bond is enhanced as shown in figure 5(b) which is presented by red. We can infer that this is the major reason why the system has the ability to store the harvested energy from the laser field.

The above analysis suggests the use of a plane capacitor model to describe the compound structure under the irradiation of laser pulses; the induced charge distribution is then modeled as two sheets of charge $\pm q$. Harvesting and storing laser irradiation energy is the procedure of charging the capacitor.

## CONCLUSION

The simulations of electron dynamics under irradiation of short laser pulses are performed by using TDDFT method in this paper. The graphene-Cu compound structure is found to be an appropriate material to harvest laser energy and convert the absorbed energy into electric power. The mechanism of conversion is investigated in our study and the effects of laser frequency as well. It is found that by subjecting the compound structure to external alternating electric fields parallel to the layers, charge transfer occurs from the copper layer to graphene sheet. At the same pulse energy, the laser with longer wavelength induces more charge transfer. The enhancement of in-plane σ-bond in graphene should be the key point to store the harvested energy for quite a long time. The basic information on graphene-cu compound structure obtained in this work is useful for developing novel energy harvesting and storing nano-scale devices such as power sources used in MEMS.

## ACKNOWLEDGEMENT

This work is supported by the National Basic Research Program of China (Grant no. 2010CB934504), strategically Leading Program of the Chinese Academy of Sciences (Grant no. XDA02040100), the National Natural Science Foundation of China (11075196, 11005142), the Shanghai Municipal Science and Technology Commission (11ZR1445200), and CAS Hundred Talents Program.

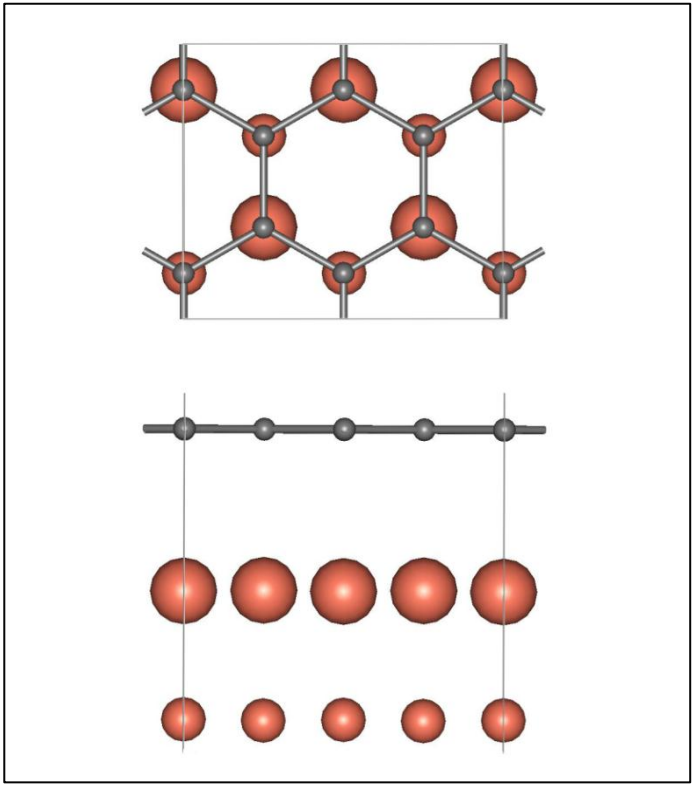

FIGURE 1 The configuration of graphene on cu (111) surface. There is one carbon atom on top of each metal atom. This geometry is corresponding to top-hcp configuration which is stable and has no higher energy than other ones.[18]

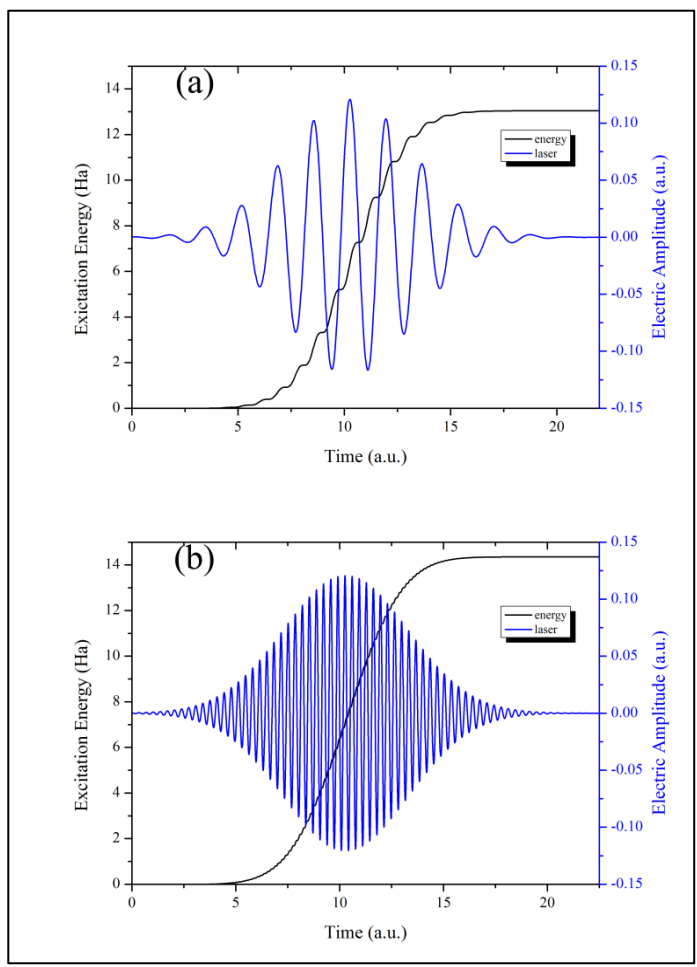

FIGURE 2 Electric field of the applied laser pulse and excitation energy as a function of time.

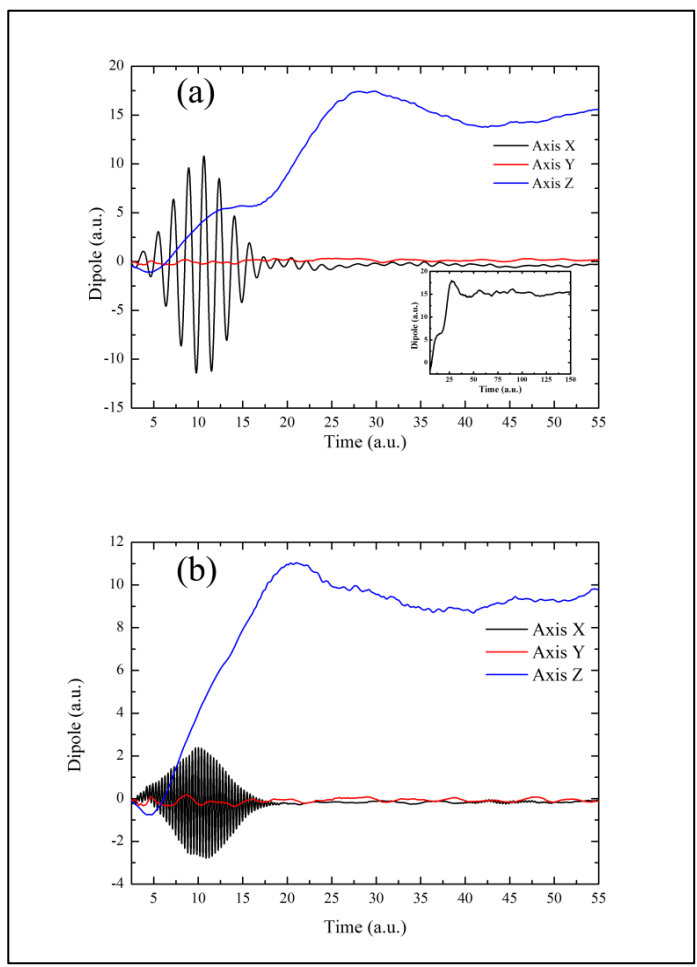

FIGURE 3 Dipoles along the 3 axes as a function of time under the irradiation of laser pulse with energy of (a) 100 eV and (b) 500 eV. The insert in (a) shows the changes of dipole along z axis for a relatively long time up to 150 a.u. (about 3.7 fs) under a lenient criterion of calculation.

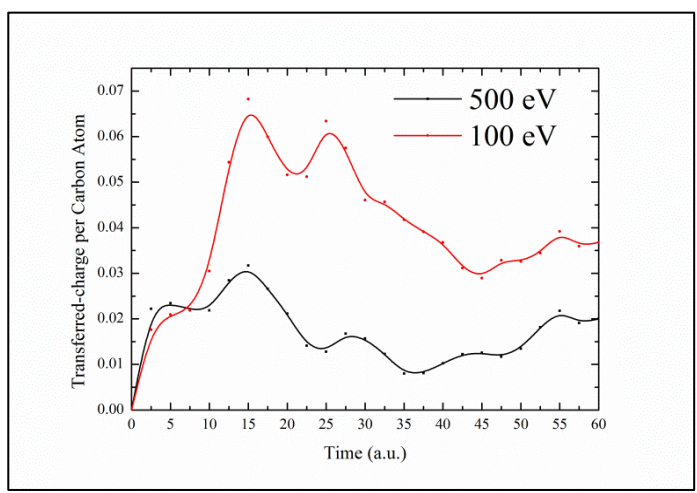

FIGURE 4 Changes of charges transferred to one carbon atom in average in a function of time. The red and black lines are corresponding to the applied laser pulses with energy of 100 eV and 500 eV respectively.

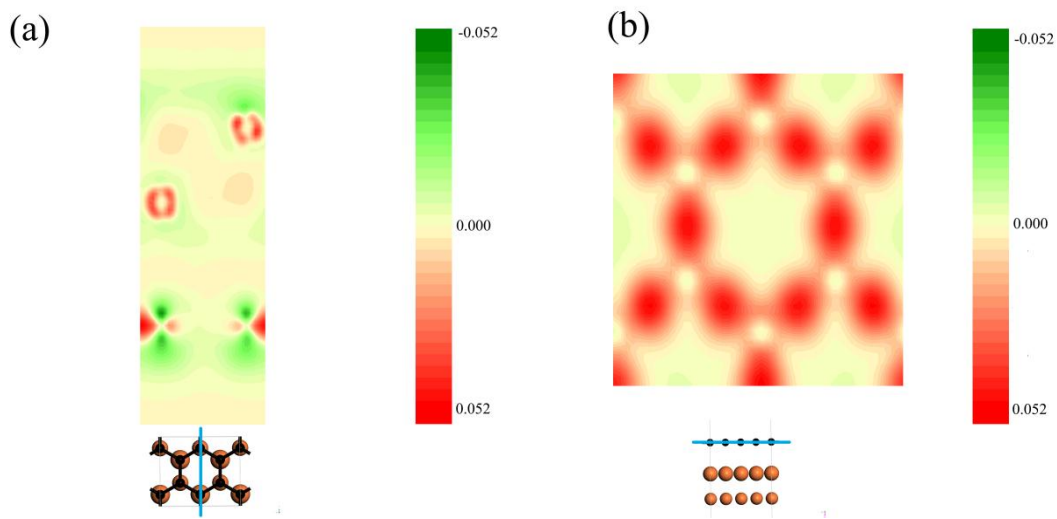

FIGURE 5 Electron density difference between the states before and after the application of laser pulse. The blue lines in the bottom indicate the position of the vertical slicing plane for plotting electron density difference. The increase of electron density is represented by reds while the green colors represent the decrease.